# Improving Students' Understanding of Lock-In Amplifiers

Seth DeVore, Alexandre Gauthier, Jeremy Levy, and Chandralekha Singh

*Department of Physics and Astronomy, University of Pittsburgh, Pittsburgh, PA 15260*

(Dated: September 1, 2015)

**Abstract:** The lock-in amplifier is a versatile instrument frequently used in physics research. However, many students struggle with the basic operating principles of a lock-in amplifier which can lead to a variety of difficulties. To improve students' understanding, we have been developing and evaluating a research-based tutorial which makes use of a computer simulation of a lock-in amplifier. The tutorial allows students to realize their conceptual difficulties by comparing their predictions about how a lock-in amplifier will behave with the outcome of simulated experiments. After being confronted with discrepancies between their predictions and the lock-in amplifier results, the tutorial targets the common difficulties that students have and leads them to develop a deeper understanding of the principles of lock-in amplification.

## I.     INTRODUCTION

The lock-in amplifier ("lock-in") is an instrument used extensively in laboratory research, especially in condensed matter physics.[1-5] In general, beginning researchers receive little to no formal training in the use of a lock-in; often, their initial exposure is superficial and successful, leading them to believe that they have mastery over its function and use. Improper or inefficient use of the lock-in, and misinterpretation of data obtained from such use, is unfortunately quite common. Additionally, this lack of understanding can result in the student's inability to troubleshoot and modify the experimental setup when faced with anomalous results.

Computer and web-based tools are becoming increasingly commonplace to aid in learning across many science and engineering fields.[6-17] For such tools to be maximally effective, it is important that they be developed using a research-based approach to ensure that they suit both the level and the prior experience of the students who will use them.[18,19] We have developed a lock-in amplifier tutorial to ease the transition of students who are just beginning their research in a laboratory setting, and to also provide a firmer foundation for those who have already used lock-ins in their research. The lock-in tutorial focuses on helping students build a robust understanding of the fundamental operation of a lock-in, and aids students in developing an intuitive feel for many of the possible situations that they may encounter in their experiments. It is integrated deeply with a computer-based model of a lock-in that is used for teaching concepts as well as testing student understanding.

Our goal in developing this research-based tutorial was to develop tools that can instill an intuitive understanding of the basics of the lock-in functions, so that students who use lock-ins in their research understand more deeply how the input signals and lock-in parameters affect the output. By merging conceptual and mathematical aspects of the instrument, the tutorial strives to help students learn the relationship between the input parameters and expected outputs so that they are able to troubleshoot unexpected outputs in their lab work.

In the following sections we begin with a description of an idealized lock-in in which we will go into some detail about the mathematical foundation for lock-in amplification. The structure of the lock-in amplifier tutorial is then described, followed by an examination of the most prevalent student difficulties identified in our initial investigation. Finally we summarize the results of the pretest and posttest scores to gauge the effectiveness of the tutorial.

## II. THE IDEAL LOCK-IN AMPLIFIER

Throughout this paper, as in the tutorial, we work with an idealized version of a lock-in amplifier. We assume that the signal of interest is centered on a frequency $f_S$ which is present in the input signal. In general it will not be a pure frequency since the amplitude can change—amplitude modulation leads to sidebands that surround the central frequency. In the case where there is no amplitude modulation introduced into the signal we will treat this frequency as a pure frequency. Tutorial questions involve both single-frequency sources as well as amplitude-modulated single-frequency sources. To separate the signal of interest from unwanted noise, a reference is defined. The reference has unit amplitude, and is dimensionless (for convenience). The single-frequency input signal is first pre-amplified by a factor $g$, to give

$$V_I = gA_S\cos(2\pi f_S t + \varphi) \tag{1}$$

This amplified signal is then multiplied (or "mixed") by a reference for the $x$ and $y$ channels of the lock-in:

$$v_{RX} = \cos(2\pi f_R t) \tag{2}$$

$$v_{RY} = \sin(2\pi f_R t) \tag{3}$$

to form the "unfiltered" $x$ and $y$-channel outputs of the lock-in:

$$V_{MX}(t) = V_I(t)v_{RX}(t) \tag{4}$$

$$V_{MY}(t) = V_I(t)v_{RY}(t) \tag{5}$$

Here, $\varphi$ is the phase of the input signal of frequency $f_S$ with respect to the reference signal, and $A_S$ is the amplitude of the input signal with frequency $f_S$. To understand the effect of the mixer, we rely on two trigonometric identities:

$$\cos(a)\cos(b) = {}^1/_2\,[\cos(a+b) + \cos(a-b)] \tag{6}$$

$$\cos(a)\sin(b) = {}^1/_2\,[\sin(a+b) - \sin(a-b)]. \tag{7}$$

Application of these identities yields:

$$V_{MX} = V_I v_{RX} = \tfrac{1}{2}gA_S[\cos(2\pi(f_S - f_R)t + \varphi) + \cos(2\pi(f_S + f_R)t + \varphi)] \tag{8}$$

$$V_{MY} = V_I v_{RY} = \tfrac{1}{2}gA_S[\sin(2\pi(f_S - f_R)t + \varphi) - \sin(2\pi(f_S + f_R)t + \varphi)] \tag{9}$$

In most experimental situations, the signal is close in frequency to the reference: $f_S - f_R \ll f_R$ and $f_S + f_R \approx 2f_R$. For the case $f_S = f_R$, we have $f_S - f_R = 0$ and $f_S + f_R = 2f_R$, and the unfiltered *x*-channel and *y*-channel outputs contain a rapidly oscillating term superimposed on a time-independent one:

$$V_{MX} = V_I v_{RX} = \tfrac{1}{2}gA_S[\cos(\varphi) + \cos(2\pi(2f_R)t + \varphi)] \tag{10}$$

$$V_{MY} = V_I v_{RY} = \tfrac{1}{2}gA_S[\sin(\varphi) - \sin(2\pi(2f_R)t + \varphi)] \tag{11}$$

Finally, $V_{MX}$ and $V_{MY}$ are each fed through a low-pass filter with a "time constant" $\tau = 1/(2\pi f_c)$ where $f_c$ is the "cutoff" or "corner" frequency of the filter; the filter "rolloff" is most commonly chosen to be one of four values (6 dB/octave, 12 dB/octave, 18 dB/octave, and 24 dB/octave). The values selected for both the time constant and the rolloff should be chosen carefully based upon the nature of the experiment. As a "rule of thumb", the $6n$ dB/octave filter preserves signals with frequency $f \ll f_c$, while attenuating signals with $f \gg f_c$ according to a power law $f^{-n}$ for (e.g., $\propto f^{-2}$ for 12 dB/octave filters). The resulting filtered outputs are defined as $V_{OutX}$ and $V_{OutY}$ in the tutorial. In the idealized version of the case where $f_R = f_S$, the time constant should be selected such that the low-pass filter should attenuate the second-harmonic $(2f_R)$ term from both $V_{MX}$ and $V_{MY}$ resulting in a time-independent output signal. The relationship between $V_{OutX}$ and $V_{OutY}$ and the magnitude and phase of the input signal as initially defined is given by the familiar trigonometric identities:

$$V_{OutX} = \tfrac{1}{2}gA_S \cos\varphi \tag{11}$$

$$V_{OutY} = \tfrac{1}{2}gA_S \sin\varphi \tag{12}$$

or

$$A_S = (2/g)\sqrt{V_{OutX}^2 + V_{OutY}^2} \tag{13}$$

$$\varphi = \tan^{-1}(V_{OutY}/V_{OutX}) \tag{14}$$

The most common use of a lock-in in laboratory research is to measure small signals that are synchronous with an external reference, often in the presence of large background signals (or "noise"). For this class of measurements, the reference frequency is set equal to the signal frequency ($f_R = f_S$). However, there are applications in which the two frequencies would not be the same; alternately, when changing a parameter that affects the signal, one is effectively modulating the signal in time. This type of "amplitude modulation" will produce sidebands around $f_S$ which must be passed with acceptably low attenuation, while making sure that the signal is not 'flooded' with background noise. While the lock-in is most commonly used in the ideal case ($f_R = f_S$ with the $2f_R$ strongly attenuated) with no other superimposed signals, there are many instances in which unwanted signals can interfere with measurement. Power line noise (i.e. the 50 Hz or 60 Hz frequency introduced by ac electrical power) can often superimpose large sinusoidal signals over the desired measurements. An understanding of the most common uses (and abuses) of lock-in amplifiers represent the primary focus of the lock-in tutorial

### III. METHODOLOGY

Before developing the lock-in tutorial, we interviewed both professors and graduate students who had prior experience using lock-ins in their research, in order to gauge what the most common uses of the lock-in are and what the most common difficulties are for new users. Alongside the development of the initial version of the tutorial, we developed a lock-in simulation which was built into the structure of the tutorial. The simulation was built with the intention of allowing students to develop an understanding of the lock-in by giving them the opportunity to experience the device first hand. We then interviewed graduate students while they made use of the most up to date version of the tutorial using a think-aloud protocol to better understand their difficulties and to fine-tune the tutorial as well as its associated pretest, posttest, and simulation. The tutorial (along with the pretest and posttest) was iteratively refined over 30 times, based upon feedback from graduate students and professors.

As the tutorial and its supplementary material underwent this process of final revision and fine-tuning based on think-aloud interview feedback, it was administered to 21 additional physics graduate students who had not been involved in the development phase of the tutorial. These students ranged in experience from those who had been introduced to the basics of the lock-in but had never made use of one themselves, through those with extensive experience with the lock-in and who either concurrently used a lock-in for their research or had made extensive use of one in the past. These students were administered the pretest and posttest before and after the tutorial in order to assess its effectiveness. To adequately evaluate the effectiveness of this tutorial despite the modifications that were made to the pretest and posttest a generalized grading rubric was developed that is capable of grading questions with all of the possible outcomes that could result when solving each of these problems. To this end, three researchers jointly deliberated a series of seven rubrics that could be utilized in the scoring of all problems present in all versions of the pretest and posttest. This resulted in the agreement of the researchers on the rubrics for scoring performance on pretest and posttest questions.[20] A more complete description of how the tutorial was developed as well as an in-depth discussion of the relevant rubrics will be given in a separate publication.

## IV. TUTORIAL STRUCTURE

The tutorial begins with the pretest. This takes the form of a short quiz comprised of "puzzles" related to the operation of a lock-in for various concrete situations. Students are provided with a PowerPoint (or web-based) presentation that contains supplementary information to guide the students as they work through this pretest. For each of the lock-in puzzles, the student is presented with a screenshot of the simulation interface, with some of the inputs or outputs hidden, and the student is asked to provide (or sketch) the missing information. The supplementary information provided contains a description of the simulation interface showing students how each of the controls modify the input signal, reference signal and low-pass filter.

After students have completed this pretest they work through the tutorial. The tutorial begins with a brief comparative analysis of several other measurement devices (the voltmeter, oscilloscope and spectral analyzer) and the lock-in. This serves to motivate the value of the lock-in for making accurate measurements of the amplitude of a specific frequency within a given signal. This leads into an in depth examination of the dual-channel lock-in. This section contains two short narrated video followed by a series of slides that provide a detailed explanation of a diagram of the dual-channel lock-in, including the basic function of the primary components.

After this basic treatment of the lock-in, students work with the simulation developed alongside the tutorial. The simulation allows the students to manipulate all of the settings commonly found on a lock-in as well as modify the characteristics of a variety of simulated input signals. The lock-in settings that the student is able to manipulate include the reference frequency, time constant and rolloff of the low-pass filter. The user can specify the frequency, amplitude and phase of the primary input signal. A sinusoidal amplitude modulation can also be introduced to the primary input signal by specifying the amplitude and frequency of the modulation. A secondary frequency can also be added to the input signal to simulate sources of interference. Finally "white" (frequency-independent) noise can be introduced into the input signal.

The interactive portion of the tutorial is prefaced by a few examples that are similar in format to the questions that follow. Students are guided through an example of how to predict the output signal of the lock-in by first applying the equations derived earlier in the mathematical treatment of the mixer and then applying the rules of thumb regarding the low-pass filter. The student then works through a series of problems designed to be used with the simulation similar to the one in Fig. 1. For each problem, the student is asked to predict the output signal when provided with a specific configuration of input signal and lock-in settings. After they have sketched their predictions, students enter the parameters into the lock-in simulator and compare the result of the simulation with their prediction. For each lock-in puzzle, two complementary explanations are given. One focuses on the mathematical signal processing, while the other attempts to provide a more conceptually intuitive explanation of the solution.

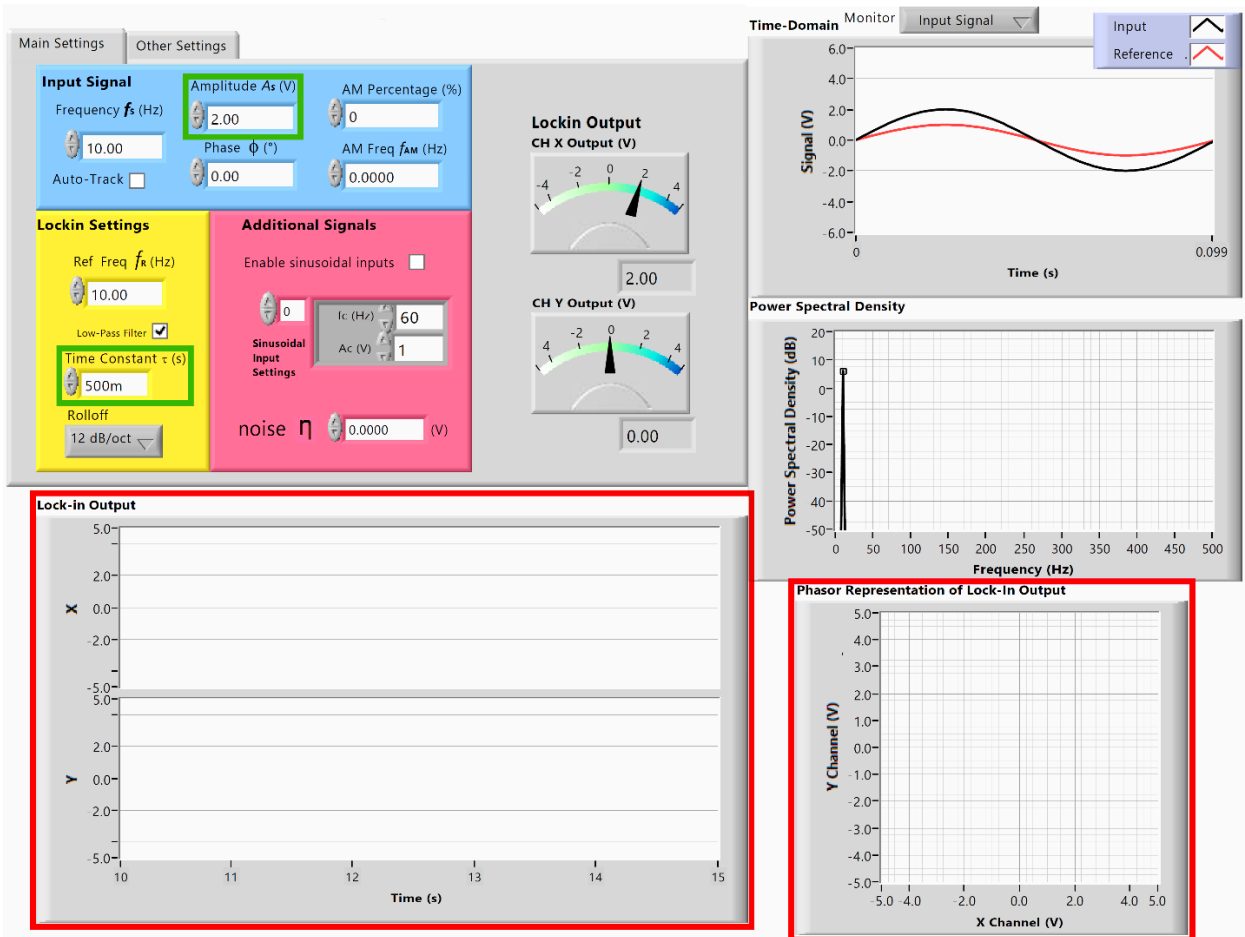

Fig. 1. Structure of a typical lock-in puzzle. The diagram shows the front panel of the lock-in simulator in a given state. Red bounding boxes indicate information that the student is expected to sketch or provide, consistent with the lock-in settings and/or output. Green bounding boxes highlight settings that were changed from a prior simulation.

The posttest is structured similarly to the pretest, consisting of a short quiz comprised of different puzzles that cover the same types of situations as in the pretest. Students are again allowed access to the same supplementary material for the posttest as in the pretest. The most up-to-date version of the tutorial, simulation and the associated pretest and posttest are available for download on ComPADRE.[21]

## V. STUDENT DIFFICULTIES

The interviews with students revealed a lack of coherent understanding of the fundamentals of a lock-in. For example, students often had a fuzzy conception of what the mixers in a lock-in do. Even for the most common and basic experimental arrangements, many students demonstrated only a superficial understanding of how a lock-in functions. The range and prevalence of these difficulties confirms that students are often using the lock-in as a "black box". Below we summarize the most common difficulties revealed by these students prior to working with the lock-in tutorial:

- **Difficulty in determining the most appropriate corner frequency (or time constant):** Interviewed students had great difficulty with the fact that the frequencies that will make it into the output signal can be estimated by making use of the time constant, $\tau$.

- **Difficulty understanding the function of the lock-in mixer:** Students often believe that the lock-in's output should have a frequency equal to that of the signal frequency. These students believe that the lock-in is providing an output similar to that which would be shown by an oscilloscope measuring the input signal.

- **Difficulty understanding the heterodyne case: $f_S \neq f_R$** : Students commonly believe a lock-in will completely filter out every signal not equal to the reference frequency. Students also commonly believe that the lock-in will produce a DC output even under the condition $f_S \neq f_R$.

- **Difficulty with the case in which $f_S \neq f_R$ and the two frequencies are out of phase ($\varphi \neq 0$):** A surprising difficulty that students expressed was the belief that, for the case where $\varphi \neq 0$ and $f_S \neq f_R$, the amplitude of the x-channel is $A_S \cos(\varphi)$ and the amplitude of the y-channel is $A_S \cos(\varphi)$. Students also showed weaker performance on problems in which $f_S = f_R$ and $\varphi \neq 0$, though this added difficulty did not result in any common incorrect answers.

- **Difficulty with the case in which $f_S = f_R$ and the $2f_R$ signal is not strongly attenuated:** Another situation that caused a considerable amount of confusion among students are situations in which $f_S = f_R$ and the $2f_R$ signal appears in the output because it is not strongly attenuated by the low-pass filter.

- **Difficulty with amplitude modulation of the input signal:** Students commonly believe that amplitude modulation necessarily affects both the x-channel and y-channel outputs equally. Students also believe that amplitude modulation should not be affected by the low-pass filter or completely forget to consider the effects of the low-pass filter on this signal.

- **Difficulty with multiple frequencies present in the input signal:** One final aspect of the lock-in that practically all students had difficulty with are cases in which multiple frequencies are present in the input signal.

Table I. Summary of the average score, standard deviation and number of instances of predicting input and predicting output questions as well as the total average score in both the pretest and posttest.

| | Average Pretest Score | Pretest Standard Deviation | Number of Instances in Pretest | Average Posttest Score | Posttest Standard Deviation | Number of Instances in Posttest | p-value |
|---|---|---|---|---|---|---|---|
| Predicting Output | 38.1% | 42.6% | 135 | 87.7% | 27.7% | 151 | <0.001 |
| Predicting Input | 42.5% | 42.1% | 55 | 80.9% | 30.5% | 55 | <0.001 |
| Total Score | 39.4% | 42.5% | 190 | 85.9% | 28.7% | 206 | <0.001 |

## VI. IMPACT OF THE TUTORIAL ON STUDENT UNDERSTANDING

In addition to discussions with faculty members, the difficulties summarized above were determined by examining student's interactions with the pretest as well as the tutorial. To evaluate the effectiveness of the tutorial, a pretest and posttest are given to all students who make use of the tutorial. In order to examine how well the tutorial addresses these difficulties, six matched pairs of lock-in puzzles were developed into pretest/posttest questions. Table I summarizes the average pre-test and post-test scores from 21 physics graduate students who took the survey. The questions are sorted into those for which the students had to predict the output, and those which the students had to predict the input. Students initially showed a limited ability to predict the output signal for the array of cases covered in the pretest. Student performance on these questions before exposure to the tutorial is roughly 40% across all pretest questions that asked the student to predict either the input or output. This average improved to over 80% across all corresponding posttest questions. Improved student performance on either of these two types of problems is likely to correlate with increased student ability to troubleshoot difficulties that may arise when making use of the lock-in.

## VII. SUMMARY

We find that a majority of physics graduate students who use lock-ins for their experimental research share many common difficulties related to the basic operation of the instrument. The difficulty lies not with the (high-school level) mathematics, but rather with the manner in which they are introduced to its use and whether students develop an integrated conceptual and quantitative understanding. The goal of the lock-in tutorial is to prepare students for both basic and more advanced (and trouble-free) usage of lock-in before they encounter these situations in real laboratory settings. The tool that we have developed and evaluated helps students of different backgrounds, with a variety of prior preparations, learn the basics of how this instrument operates and also helps students make connections with the underlying mathematics that describes the operations of its major components. The tutorial also develops the student's ability to predict output signals and input signals in a variety of cases likely to be encountered by students in laboratory setting.

Examination of the average student scores on the pretests and posttests showed considerable improvement for all cases discussed in the tutorial. While there are many topics covered in this tutorial, it is straightforward to add additional topics (e.g., the treatment of broadband noise, or the output response to sudden signal amplitude changes) by coupling a brief introduction with associated lock-in puzzles.

**ACKNOWLEDGEMENTS**

This work was supported by the National Science Foundation, award NSF-1124131.